\documentclass[a4paper,twocolumn,11pt]{quantumarticle}
\pdfoutput=1

\usepackage[utf8]{inputenc}
\usepackage[english]{babel}
\usepackage[T1]{fontenc}
\usepackage{amsmath}
\usepackage{hyperref}
\usepackage[numbers,sort&compress]{natbib}
\usepackage{tikz}
\usepackage{lipsum}

\begin{document}
	\title{Ordering Matters: Structure, Accuracy and Gate Cost in Second-Order Suzuki Product Formulas}
	
	\author{Matthew A Lane}
	\affiliation{Physics \& Astronomy Department, University College London}
	\orcid{0000-0002-3222-1348}
	
	\author{Dan E Browne}
	\affiliation{Physics \& Astronomy Department, University College London}
	\orcid{0000-0003-3001-158X}
	
	\maketitle
	
	\begin{abstract}
		Product formula methods, particularly the second-order Suzuki decomposition, are an important tool for simulating quantum dynamics on quantum computers due to their simplicity and unitarity preservation.
		While higher-order schemes have been extensively studied, the landscape of second-order decompositions remains poorly understood in practice.
		We explore how term ordering and recursive application of the Suzuki formula generate a broad family of approximants beyond standard Strang splitting, introducing a hybrid heuristic that minimizes local error bounds and a fractional approach with tunable sequence length.
		The hybrid method consistently selects the longest possible decomposition, achieving the lowest error but at the cost of exponential gate overhead, while fractional decompositions often match or exceed this performance with far fewer gates, enabling offline selection of near-optimal approximants for practical quantum simulation.
		This offers a simple, compiler-accessible heuristic for balancing accuracy and cost, and highlights an underexplored region of decomposition space where many low-cost approximants may achieve high accuracy without global optimization.
		Finally, we show that in the presence of depolarising noise, fractional decompositions become advantageous as systems approach fault-tolerant error rates, providing a practical path for balancing noise resistance and simulation accuracy.
	\end{abstract}
	
	\section{Introduction}\label{sec:introduction}
	Hamiltonian simulation is a central task in quantum computation, with the goal of approximating the time evolution operator $U(t) = e^{-i Ht}$ for a time-independent Hamiltonian $H$ of $L$ qubits. 
	In many systems, $H$ can be written as a sum of non-commuting terms $H = \sum_{j=1}^m H_j$, where each $H_j$ is a $k$-local term which acts non-trivially on $k$ qubits.
	
	A quantum simulator can approximate the full evolution $e^{-i Ht}$ by a sequence of unitary operators of the form $e^{-i Ht/r}$.
	As long as the timestep $\delta t=t/r$ is sufficiently small, achieved by choosing a suitably large $r$, it becomes possible to leverage product formula (PF)-based approaches to rewrite the sum $\sum_{j=1}^m H_j$ in a way that enables efficient simulation on a quantum computer as a series of elementary simulation steps.
	That is, each $e^{-i H_jt/r}$ can be implemented with quantum gates, and the full $e^{-i Ht}$ can be approximated by a sequence of such unitaries \cite{trotter1958approximation,lloyd1996universal}. 
	The resulting decomposition takes the form of an ordered product of unitary operators, naturally preserving unitarily to generate algorithms which are guaranteed to be unconditionally stable \cite{smith1985numerical,de1987product}.
	
	The origin of PFs in mathematics is most often attributed to S. Lie's work on algorithmic convergence for solving ordinary differential equations and the convergence of special integrators \cite{lie1880theorie,chorin1978product}.
	For matrices $A$ and $B$, the differential equation $dx/dt=Ax+Bx$ can be solved via the propagator $K_{\Delta t}=e^{A\Delta t}e^{B\Delta t}$, which leads to Lie's 1875 formula:
	\begin{align}\label{eq:lie-trotter}
		e^{A+B}=\lim_{r\rightarrow\infty} \left(e^{A/r}e^{B/r}\right)^r.
	\end{align}
	This expression is commonly referred to as the Lie-Trotter formula, or Trotterization \cite{trotter1958approximation} in physics.
	For the Hamiltonian $H$, the simulation works by evolving the system locally over small discretized timesteps to form the approximation $e^{-iHt}\approx(e^{-iH_1\delta t}e^{-iH_2\delta t}\ldots e^{-iH_m\delta t})^r$.
	Each of the local time-evolution operators $e^{-iH_1\delta t}\ldots e^{-iH_m\delta t}$ is applied in sequence, and the full cycle is repeated $r$ times. 
	For this reason, $r$ is referred to as the \textit{Trotter number}, and many quantum systems in condensed matter physics have been classically simulated this way with the aid of Monte Carlo techniques \cite{binder1993monte,suzuki1993quantum}. 
	Each overall timestep approximates $e^{-iH\delta t}$, introducing an $O(\delta t^2)$ error per timestep.
	This error accumulates at most linearly in $r$ as the sequence is repeated $r$ times, so the total error scales as $O(t^2/r)$.
	
	Considerable effort has gone into tightening the bounds on the approximation error of PF decompositions \cite{berry2007efficient,hadfield2018divide,low2019well,childs2021theory,layden2022first}, going well beyond the early estimates based on truncating the Baker-Campbell-Hausdorff (BCH) expansion.
	Despite this progress, the Lie-Trotter decomposition remains the most widely used PF approach, even though it is only a first-order method, accurate only to first order in the BCH expansion.
	Higher-order decomposition schemes, in which each $e^{-iH_)j\delta t}$ appears multiple times within a single timestep, have also been developed.
	Among these, the second-order Trotter-Suzuki decomposition (referred to simply as Suzuki) is notable because it takes advantage of symmetry properties within a timestep to improve accuracy and is central to the present work \cite{suzuki1985decomposition,suzuki1991general}.
	
	The simplest form of the Suzuki decomposition, analogous to the Lie-Trotter formula, is
	\begin{align}\label{eq:simple_suzuki} 
		e^{A+B} = \lim_{r \rightarrow \infty} \left(e^{A/2r} e^{B/r} e^{A/2r}\right)^r, 
	\end{align} 
	and is commonly referred to as symmetric Strang splitting \cite{strang1968construction}. 
	This is an example of a split-operator method used to solve differential equations numerically \cite{glowinski2017splitting}.
	For a Hamiltonian expressed as a sum of many terms, the simplest and most common Suzuki decomposition is given by
	\begin{align}
		e^{-iH\delta t} &\approx \left( \prod_{j=1}^{m} e^{-iH_j \delta t / 2} \right) \left( \prod_{j=m}^{1} e^{-iH_j \delta t / 2} \right) \nonumber \\
		&\equiv S_2(\lambda), \label{eq:suzuki_step} 
	\end{align} 
	where $\lambda = -i \delta t$. 
	Note that we assume the convention that matrix products always multiply from the left, and that the index ordering is reversed in the second product relative to the first.
	
	The approximation error is reduced compared to Trotterization by symmetrizing the application of terms to partially cancel contributions from their non-commutativity in the BCH expansion; this is readily seen by examining the second-order Taylor expansion, with each timestep introducing an $O(\delta t^3)$ error, leading to a global error that scales as $O(t^3 / r^2)$.
	However, Eq. \eqref{eq:suzuki_step} is only one possible decomposition for a given Hamiltonian; a variety of approximants with similar theoretical error bounds exist \cite{hatano2005finding}, all arising from Eq. \eqref{eq:simple_suzuki}.
	These equivalent decompositions differ in the number of times each $e^{-iH_j \delta t}$ appears in the sequence, their relative positions within the product, and the numerical prefactors in the exponents, each of which varies between constructions.
	
	Higher-order Suzuki PFs can be generated via a recurrence relation derived by Suzuki \cite{suzuki1991general}: \begin{align}
		S_{2k}(\lambda) &= S_{2k-2}(p_k \lambda)^2 S_{2k-2}((1 - 4p_k)\lambda) \nonumber \\
		&\qquad S_{2k-2}(p_k \lambda)^2, \label{eq:suzuki_recursion} 
	\end{align} 
	where $S_{2k-2}$ is the Suzuki approximation of order $2k-2$, and the coefficient $p_k$ is a timestep reweighting factor given by
	\begin{align}\label{eq:suzuki_weights} 
		p_k = \left(4 - 4^{1/(2k - 1)}\right)^{-1}. 
	\end{align}
	Note that $S_2$ in Eq. \eqref{eq:suzuki_step} is the lowest-order Suzuki decomposition from which all higher-order PFs are subsequently constructed using Eq. \eqref{eq:suzuki_recursion}; this work is interested in making a choice for the structure of $S_2$ that is better than Eq. \eqref{eq:suzuki_step} in terms of decomposition error, reducing the fidelity loss between the exact density matrix and the density matrix evolved by the decomposition. 
	
	Expressions Eqs. \eqref{eq:suzuki_recursion}-\eqref{eq:suzuki_weights} were constructed \cite{ruth1983canonical,suzuki1991general} to systematically cancel higher-order terms in the Taylor expansion such that $S_{2k} = S_{2k-1}$, meaning that only even-order decompositions are used in practice.
	Any higher-order approximation generated via this recursion ultimately consists of a product of second-order approximations, with each exponent scaled by a set of phase factors determined by the reweighting coefficients $p_k$.
	As a result, there are only two key parameters that govern the approximation: the order $k$, which determines the structure of the decomposition, and the number of timesteps $r$.
	In many cases, these parameters are in competition: extra quantum gates can either be allocated to increasing the order of the decomposition or to reducing the timestep size.
	While the error bounds associated with each choice are well understood, the actual simulation error observed in practice is not.
	It is also worth emphasizing that Eqs. \eqref{eq:suzuki_recursion} and \eqref{eq:suzuki_weights} were constructed so that the PFs generated would be accurate to a given order in the Taylor expansion, but the set of values $p_k$ for a given $k$ represent just one possible choice and are not necessarily optimal. 
	Therefore, Eqs. \eqref{eq:suzuki_recursion} and \eqref{eq:suzuki_weights} should not be viewed as defining the theoretical best decomposition.
	
	Numerical studies of Suzuki decomposition error have typically focused on higher-order schemes, which necessarily result in increasingly large gate counts since they are built from products of second-order decompositions \cite{thalhammer2008high,thalhammer2012convergence,childs2018toward,jones2019optimising}.
	Surprisingly little attention, however, has been given to optimizing the second-order Suzuki decomposition despite the fact that, as with higher-order cases, there exist many mathematically equivalent expressions with the same theoretical error bounds that may perform differently in practice.
	For the Lie-Trotter formula, the arbitrary ordering of exponentials $e^{-iH_j \delta t}$—a consequence of the arbitrary ordering of terms in the Hamiltonian—can be exploited to reduce gate count and circuit depth \cite{lao20222qan}.
	Similar efforts to minimize parallel depth have been made in quantum chemistry simulations using Suzuki decompositions \cite{hastings2014improving}.
	It has also been shown that randomly choosing certain approximant structures can lead to improved algorithms with reduced decomposition error \cite{zhang2012randomized}, and that significant improvements in error bounds can arise when the order of terms is randomly permuted \cite{childs2019faster}.
	Nevertheless, no systematic heuristic currently exists for choosing the term ordering in Suzuki decompositions more broadly, nor for identifying those with particularly beneficial properties.
	
	This work begins to address some of these points by identifying extreme examples of the second-order Suzuki decomposition $S_2$ in terms of maximal and minimal gate count, demonstrating that the performance of different approximants with the same error bound can vary by more than an order of magnitude.
	We highlight one specific choice that consistently performs as well as others with significantly higher gate counts, potentially serving as a simple heuristic for designing future decompositions, being an improved building block for higher-order PFs generated via the Suzuki recursion relation.
	In addition, we consider the effect of depolarising noise and identify a crossover regime where, as the depolarisation probability decreases into the approximate fault-tolerant range, it becomes advantageous to favour the longer sequence length decompositions developed here, which exhibit lower approximation error, over pre-existing shorter decompositions that are less accurate.
	
	Finally, it should be noted that alternative methods for simulating quantum dynamics which do not rely on product formulas have been developed, offering better asymptotic performance than Trotter-based approaches \cite{berry2014exponential,berry2015hamiltonian,low2017optimal}.
	However, recent advances in tightening the error bounds associated with PFs \cite{childs2021theory} have helped explain the better-than-expected performance of more sophisticated Trotterization techniques, even when judged against earlier, more conservative bounds \cite{hadfield2018divide,childs2019faster,ouyang2020compilation,layden2022first}.
	
	\section{Variety of Second-Order Suzuki Decompositions}\label{sec:variety}
	\subsection{Sequential Application of the Suzuki formula}
	For a Hamiltonian consisting of more than two terms, Eq. \eqref{eq:simple_suzuki} does not by itself provide a complete prescription of the decomposition.
	In the case of the three-term Hamiltonian $H=H_1+H_2+H_3$, Eq. \eqref{eq:simple_suzuki} must be applied at least twice, each time treating one of the summands as either $A$ or $B$ in the formula.
	
	For example, the first application of Eq. \eqref{eq:simple_suzuki}, starting with $H_1$, gives rise to both of  the following: 
	\begin{align}
		&e^{-iH\delta t} \approx e^{-iH_1\delta t/2} e^{-i(H_2+H_3)\delta t} e^{-iH_1\delta t/2} \\
		&e^{-iH\delta t} \approx e^{-i(H_2+H_3)\delta t/2} e^{-iH_1\delta t} e^{-i(H_2+H_3)\delta t/2}. \label{eq:3term_2}
	\end{align}
	Eq. \eqref{eq:simple_suzuki} can then be applied again to the remaining sum $H_2+H_3$, for instance by pulling out $H_2$ in the same way
	This yields two possible second-order decompositions:
	\begin{align}
		e^{-iH\delta t} &\approx e^{-iH_1\delta t/2} e^{-iH_2\delta t/2} e^{-iH_3\delta t} e^{-iH_2\delta t/2} \nonumber \\
		&\qquad e^{-iH_1\delta t/2} \label{eq:3term_min} \\
		e^{-iH\delta t} &\approx e^{-iH_2\delta t/4} e^{-iH_3\delta t/2} e^{-iH_2\delta t/4} \nonumber \\
		& e^{-iH_1\delta t} e^{-iH_2\delta t/4} e^{-iH_3\delta t/2} e^{-iH_2\delta t/4}. \label{eq:3term_max}
	\end{align}
	Eq. \eqref{eq:3term_min} is an example of Strang splitting and is typically regarded as the standard second-order Suzuki decomposition.
	However, Eq. \eqref{eq:3term_max} is equally valid, and these are not the only two possibilities.
	The ordering of $H_1$, $H_2$ and $H_3$ is arbitrary: the terms could be arranged lexicographically, by magnitude, or according to some other heuristic.
	Some evidence suggests that ordering them by magnitude may yield improved performance in some cases \cite{hadfield2018divide}. 
	Moreover, the product need not be symmetric; the left and right halves of the decomposition can differ. 
	
	The two examples, Eqs. \eqref{eq:3term_min} and \eqref{eq:3term_max}, are noteworthy because they correspond to the shortest and longest possible sequence lengths for a three-term Hamiltonian, with direct implications for circuit depth. 
	For Hamiltonians with more terms, each recursive application of Eq. \eqref{eq:simple_suzuki} involves choosing whether to place the extracted term in the duplicated $A$ (outer) positions or the single $B$ (middle) position. 
	Placing the term in $B$ results in a shorter sequence, while placing the remaining sum in the duplicated $A$ positions leads to an increase in sequence length at each step, since the formula must be applied to both instances, though never with full doubling. 
	This generates a range of decompositions: at one extreme, placing each new term in $B$ yields the minimal sequence length $2m - 1$, where $m$ is the number of terms in the Hamiltonian, equivalent to standard Strang splitting; at the other, placing the remaining sum in $A$ at each step leads to the recurrence $N_m = 2N_{m-1} + 1$ for the sequence length after $m$ applications, producing a maximum length of $2^m - 1$, the size of a full binary tree with $m$ leaves. 
	
	Having established the range of possible sequence lengths, we now consider how to choose the order in which terms are extracted from the Hamiltonian, and whether to apply a \textit{shallow} or \textit{wide} decomposition step at each stage.
	A \textit{shallow} step places the selected term in the outer ($A$) positions of Eq. \eqref{eq:simple_suzuki} and the rest of the sum in the middle ($B$) position, resulting in a shorter sequence.
	\begin{align}
		e^{\textnormal{term}+\textnormal{the rest}} \approx e^{\textnormal{term}/2} e^{\textnormal{the rest}} e^{\textnormal{term}/2} \nonumber
	\end{align}
	A \textit{wide} step places the selected term in the middle ($B$) position and the rest of the sum in the outer ($A$) positions, effectively duplicating the remainder of the sum so that it appears twice in the sequence.
	\begin{align}
		e^{\textnormal{term}+\textnormal{the rest}} \approx e^{\textnormal{the rest}/2} e^{\textnormal{term}} e^{\textnormal{the rest}/2} \nonumber
	\end{align}
	
	Both the term ordering and step type at each level influence not only the final length, but also the structure and numerical behavior of the decomposition.
	This creates a large combinatorial space: for a Hamiltonian with $m$ terms, one must choose from $m!$ orderings and $2^{m-1}$ decomposition patterns.
	Exhaustively exploring this space is infeasible even for moderate values of $m$.
	
	\subsection{Term Ordering Heuristic}\label{sec:term_ordering}
	Since the Lie-Trotter and Suzuki decompositions are sensitive to non-commutativity, placing strongly non-commuting terms adjacent to one another can amplify the approximation error, while separating them, or ordering terms by magnitude or structure, may help suppress it \cite{tranter2019ordering}.
	An alternative strategy is to use the local error bound at each application of the Suzuki formula to select the type of decomposition and term ordering.
	For the two-term Hamiltonian $A + B$, the relevant error bound is given by \cite{suzuki1985decomposition,de1987product},
	\begin{align}
		&\big\vert\big\vert e^{-i\delta t(A+B)} - e^{-i\delta tA/2}e^{-i\delta tB}e^{-i\delta tA/2} \big\vert\big\vert \nonumber \\
		&\quad\leq \delta t^3 \left( \big\vert\big\vert [A,[A,B]] \big\vert\big\vert +2 \big\vert\big\vert [B,[A,B]] \right) \big\vert\big\vert, \label{eq:step_error}
	\end{align}
	where $A$ and $B$ take the role of the term being pulled out of the sum or the rest of the sum. 
	We then choose whether to perform a shallow or wide decomposition step based on which decomposition will contribute the lowest error bound for that decomposition step, resulting in a \textit{hybrid} decomposition which switches between \textit{shallow} and \textit{wide} as appropriate, presented in Algorithm 1. 
	
	\begin{figure}[ht]
		\includegraphics[width=\linewidth]{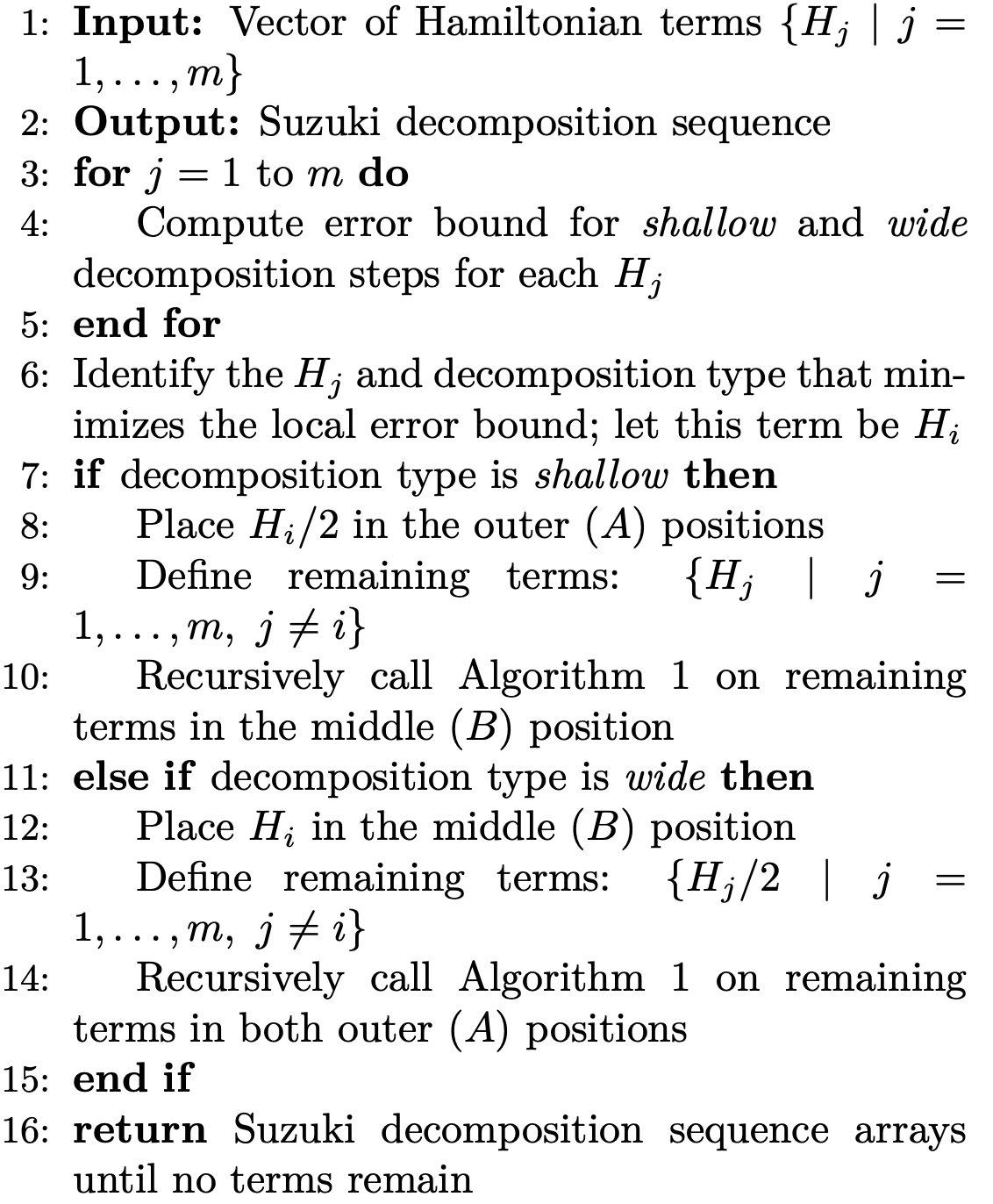}
		\captionsetup{width=\linewidth}
		\caption*{Algorithm 1: Hybrid Suzuki Decomposition \label{fig:algorithm}}
	\end{figure}
	
	
	This hybrid heuristic provides a principled way to select the decomposition structure by minimizing the leading-order local error at each recursive step. 
	However, its numerical performance, and whether it outperforms simpler strategies like fixed-order Strang splitting which always has less than or equal circuit depth, depends on the commutator structure of the Hamiltonian and the trade-off between sequence length and accumulated error.
	Crucially, the method performs only local optimization and does not yield a globally minimal decomposition. 
	Identifying a global optimum would require exploring all possible term orderings and decomposition paths which amounts to classically simulating the full quantum evolution. 
	The hybrid approach thus represents a first attempt at a tractable compromise.
	
	As a complementary strategy, we also consider a non-adaptive variant where, instead of making a locally optimal choice at each step, we fix a desired proportion of wide decomposition steps, i.e., steps where the selected term is placed in the middle position, with the remainder of the sum in the outer positions. 
	This proportion is enforced deterministically during the recursive construction of the decomposition until the target ratio is reached.
	This effectively functions as a variant of Algorithm 1 where: the term with minimal local error bound is found and pulled out of the sum using a wide step; this is repeated until the desired ratio of wide to shallow steps is achieved; after this, only shallow decomposition steps are applied.
	
	Although not optimized for performance, this approach allows us to systematically explore how the structural balance between shallow and wide steps affects the overall error and gate count.
	For the sake of brevity, we refer to this as the \textit{fractional} decomposition. 
	
	\section{Empirical Results}
	\subsection{Test Systems}
	To evaluate the performance of different decomposition strategies, we consider three test models defined on a one-dimensional chain of $L$ qubits: the transverse field Ising model (TFIM), the Heisenberg XYZ model, and a random Pauli Hamiltonian. These models were chosen as benchmarks because they exhibit non-trivial quantum dynamics, strong operator non-commutativity, and, in the generic case, non-integrability, making them well-suited for testing the accuracy and robustness of simulation algorithms.
	Note that non-commutativity will in general always arise, since we are interested in the commutativity of a summand with the remainder of the sum rather than between individual summands. 
	
	\subsubsection{1D Transverse Field Ising Model}
	The one-dimensional transverse field Ising model is a paradigmatic spin system used to study quantum phase transitions and many-body quantum dynamics. 
	It consists of a qubit chain with nearest-neighbour Ising interactions and a uniform transverse magnetic field, described by the Hamiltonian,
	\begin{align}
		H_{\textnormal{TFIM}} = -J\sum_{j=1}^{L-1} Z_j Z_{j+1} - h\sum_{j=1}^{L} X_j,
	\end{align}
	where $J$ is the interaction strength and $h$ is the transverse field strength. 
	The model is exactly solvable in the thermodynamic limit and exhibits a quantum phase transition at the critical point $h = J$, separating a ferromagnetic phase from a paramagnetic one \cite{pfeuty1970one}.
	
	\subsubsection{Heisenberg XYZ Model}
	The one-dimensional Heisenberg XYZ model describes a chain of spin-$\frac{1}{2}$ particles with anisotropic nearest-neighbour interactions along the three spatial directions. Its Hamiltonian is
	\begin{align}
		H_{\textnormal{XYZ}} &= \sum_{j=1}^{L-1} \left( J_x X_j X_{j+1} + J_y Y_j Y_{j+1} \right. \nonumber \\
		&\qquad\qquad \left. + J_z Z_j Z_{j+1} \right),
	\end{align}
	where $J_x$, $J_y$, and $J_z$ are the coupling strengths along the $x$, $y$, and $z$ directions. 
	The XYZ model generalises the XX and XXZ models and features rich dynamical behaviour and symmetry breaking depending on the anisotropy of the couplings \cite{baxter2016exactly}.
	
	\subsubsection{Random Pauli Hamiltonian}
	We also consider a random Pauli Hamiltonian composed of on-site and nearest-neighbour interaction terms drawn randomly from the Pauli basis. 
	Each on-site term is chosen independently from $\{X, Y, Z\}$ with uniform probability $1/3$, while each interaction term consists of a random pair of Pauli operators acting on adjacent qubits, e.g., $X_j Z_{j+1}$ or $Y_j Y_{j+1}$, with uniform probability $1/9$ for each of the nine non-identical pairings, while all terms have equal coefficients equal to 1. 
	Such models lack translational symmetry and exhibit highly non-uniform commutation structure, making them a useful stress test for decomposition strategies sensitive to term ordering and local error propagation. 
	They also serve as a minimal toy model for disordered or engineered quantum systems with irregular coupling geometries.
	
	\subsection{Numerical Performance}
	Unless otherwise stated, the following parameters were used in all simulations: qubit-chain of $L=5$ qubits, timestep $\delta t=0.01$ over 500 time slices, and the density matrix was initialised in the all-zero pure Z-state $|0\ldots 0\rangle\langle0\ldots 0\vert$.
	For the TFIM Hamiltonian, values of $J=1$ and $h=5$ were used, while for the Heisenberg XYZ Hamiltonian, values of $J_x=3$, $J_y=2$, and $J_z=1$ were used. 
	
	To compare matrices, we use the trace distance to quantify the distinguishability between two quantum states, with values closer to zero indicating higher similarity. 
	For density matrices $\rho_1$ and $\rho_2$, the trace distance is defined as
	\begin{align}\label{eq:trace_dist} 
		D(\rho_1,\rho_2) = \frac{1}{2} \textnormal{Tr}\left|\rho_1 - \rho_2\right|,
	\end{align}
	where $|A| = \sqrt{A^\dagger A}$ denotes the operator absolute value. 
	It has the advantage of being a true metric—symmetric, contractive under completely positive trace-preserving (CPTP) maps, and directly relates to the probability of distinguishing two states via optimal measurement. 
	Although we also computed the fidelity and Bures distance for all data presented, they exhibited the same qualitative trends; the trace distance is shown here as it consistently provided the clearest separation between decomposition strategies.
	
	\subsubsection{Extremal Gate Count vs Accuracy Tradeoff}\label{sec:tradeoff}
	One of the central challenges in PF simulation is navigating the trade-off between simulation accuracy and circuit complexity. 
	We take the number of unitary factors in the Suzuki decomposition which directly corresponds to the number of local unitaries to be implemented as a proxy for gate count. 
	This allows us to quantify how the accuracy of different decomposition strategies scales with gate overhead. 
	
\begin{figure*}[ht!]
		\includegraphics[width=\linewidth]{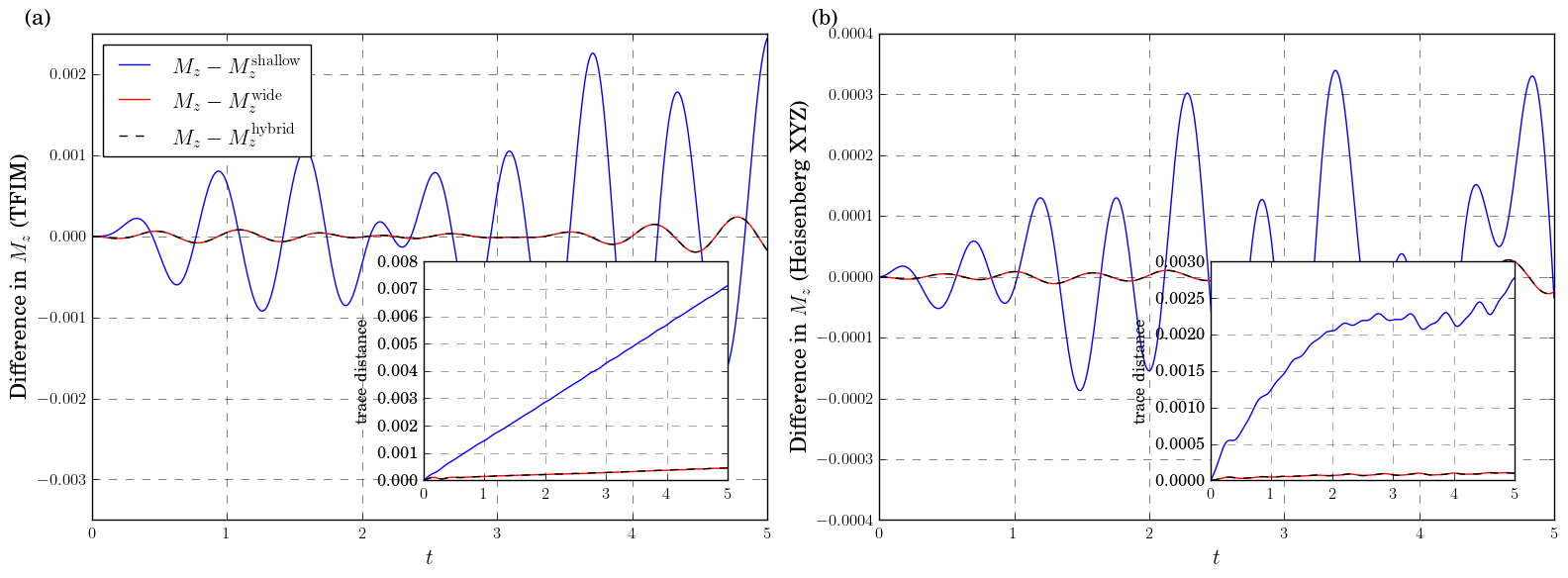}
		\captionsetup{width=\linewidth}
		\caption{(a) The TFIM Hamiltonian and (b) Heisenberg XYZ Hamiltonian. In both cases, the main panel shows the difference between the exact $X$-magnetization $M_x$ using the full unitary time-evolution operator $\textnormal{exp}\{-iH_\textnormal{TFIM}\delta t\}$ at each time slice, and the magnetization using the shallow (blue), wide (red) and hybrid (black dashed) Suzuki decompositions to generate PFs. \textit{Inset:} The trace distance for the corresponding density matrices. Note that the black dashed line and red line overlap. \label{fig:mag}}
	\end{figure*}

	While the shallow decomposition (standard Strang splitting) is optimal in terms of sequence length, it consistently exhibits the highest simulation error across the three models considered. 
	Here, sequence length refers to the number of factors appearing in the product, that is, the number of unitaries to be applied.
	Conversely, the wide decomposition, which maximizes sequence length by duplicating the remaining Hamiltonian at each recursive step, yields significantly lower error at the cost of exponentially more gates. 
	The hybrid decomposition, which minimizes the local error bound at each step, always reproduces the wide decomposition for all models considered.
	This trade-off is clearly illustrated in Fig. \ref{fig:mag}(a) for the 5-qubit transverse field Ising model (TFIM). 
	The shallow decomposition, with a sequence length of only 19, incurs approximately 15 times the magnetization error and 13 times the trace distance compared to the wide decomposition, with a sequence length of 1023. 
	This disparity grows rapidly with system size: even a 7-qubit chain ($m=14$ terms) yields sequence lengths of $2m-1=27$ (shallow) and $2^m-1=16,383$ (wide).
	
	The same pattern appears for the Heisenberg XYZ model (Fig. \ref{fig:mag}(b)), where the shallow decomposition (67 exponentials) yields magnetization errors and trace distance approximately 17 and 26 times larger, respectively, than the wide/hybrid decompositions (32,767 exponentials). 
	Despite its common use, the shallow decomposition always performs worst, even for small systems, highlighting its limitations for accurate quantum simulation.
	
	While wide decompositions consistently deliver the lowest simulation error by an order of magnitude, they do so at the cost of exponential gate overhead even for these relatively small and simple Hamiltonians. 
	This makes them impractical for large systems despite their accuracy. 
	Notably, the hybrid decomposition, designed to minimize the local error bound at each recursive step, is observed to always select the wide decomposition structure in both of these physical models. 
	This suggests that, under existing local error bounds, the wide decomposition is likely to be close to the globally superior solution that would be found by an exhaustive search through all possible decompositions. 
	
	\subsubsection{Fractional Decomposition Sweep}
	While the shallow and wide decompositions represent extremes in the accuracy–gate count trade-off, many intermediate strategies exist. 
	To explore this space, we defined a family of \textit{fractional} decompositions in which a fixed proportion of decomposition steps are forced to follow the wide structure, with the remainder defaulting to shallow. 
	This proportion is specified in advance, making it a simple and tunable parameter that avoids online optimization and can be set at compile time.
	
\begin{figure*}[ht!]
		\includegraphics[width=\linewidth]{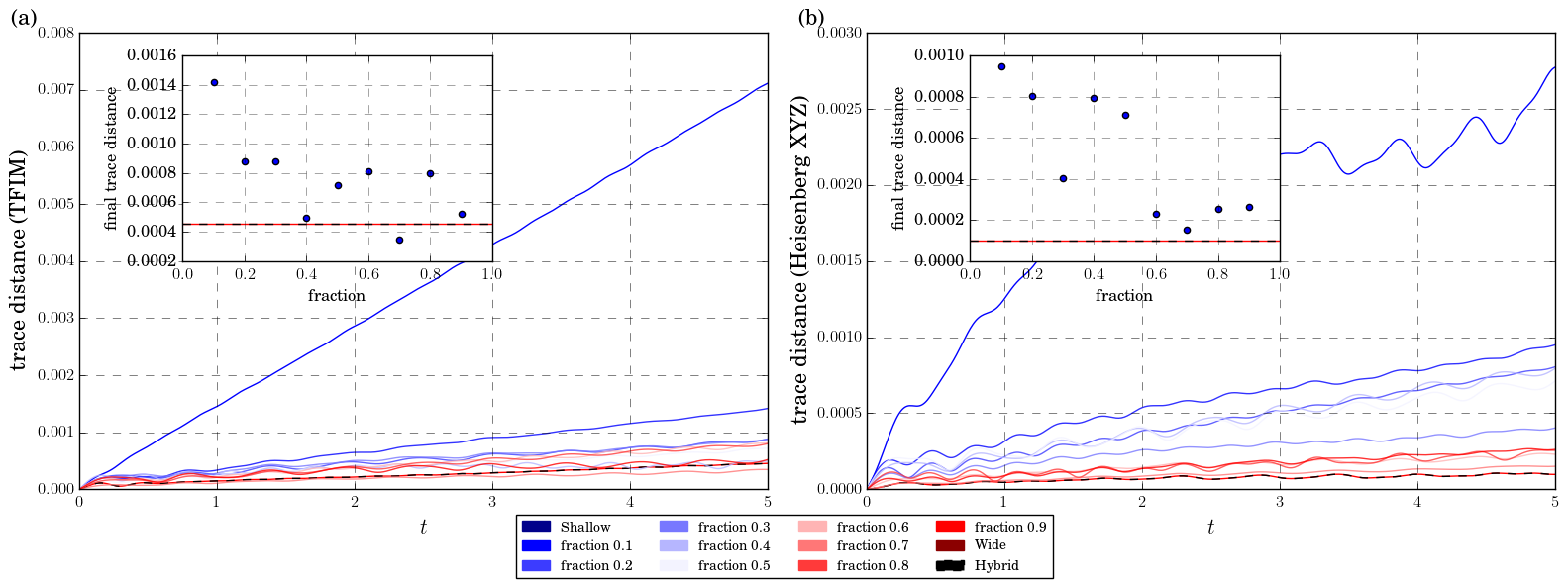}
		\captionsetup{width=\linewidth}
		\caption{The trace distance between the exact density matrix and the density matrix using a Suzuki decomposition for (a) the TFIM Hamiltonian and (b) Heisenberg XYZ Hamiltonian. The extremal shallow and wide decompositions (extremal in terms of sequence length) are shown in dark blue and dark red respectively, with the hybrid decomposition as a dashed black line, and a range fractional decompositions with wide:shallow fraction between 0.1 and 0.9. \textit{Inset:} The final values of the trace distance using the fractional decomposition as a function of the wide:shallow fraction, with the final values from the wide and hybrid decompositions shown as red and black dashed lines respectively. The shallow, wide and hybrid data is the same as in Fig. \ref{fig:mag}. \label{fig:fractional}}
	\end{figure*}

	We sweep through wide:shallow ratios from 0.1 to 0.9 in increments of 0.1 and compute the resulting trace distances to the exact evolved density matrix. 
	As shown in Fig. \ref{fig:fractional}, all fractional decompositions substantially outperform the shallow decomposition, achieving $5$–$10\times$ lower error across both models, and cluster closely around the wide and hybrid decompositions which again match. 
	For the TFIM, a 0.4 fraction achieves nearly the same trace distance as the wide decomposition but with a sequence length of 79, compared to 1023 for the wide case (shallow decomposition sequence length 35). 
	Similarly, in the Heisenberg XYZ model, a 0.7 fraction achieves comparable performance with a sequence length of 2643, far below the sequence length of the wide decomposition of 32,767 (shallow decomposition sequence length 67).
	
	These results underscore two key points. 
	First, the hybrid strategy, which always selects the wide decomposition under local error-bound minimization, does not necessarily yield the best global approximation, with fractional decompositions occasionally achieving lower trace distance despite reduced circuit depth. 
	Notably, for the TFIM, one of the fractional decompositions (0.7 fraction) achieves a lower trace distance than the wide decomposition. 
	This confirms that local minimization of the error bound at each recursive step, while useful and inexpensive, does not guarantee global optimality in terms of actual simulation error, highlighting the limitations of relying solely on local analytic bounds.
	Crucially however, since the local minimization is inexpensive, this sweep can be performed entirely offline by a compiler, allowing pre-selection of a decomposition that matches or exceeds the accuracy of the wide decomposition while using dramatically fewer gates, offering a powerful and cost-free (in terms of quantum resources) optimisation step for quantum simulation pipelines.
	
	\subsubsection{Generality and Variability across Random Pauli Models}
	To test the robustness of our findings beyond structured models, we analyze an ensemble of 100 random Pauli Hamiltonians, each defined on a 5-qubit chain with randomly selected on-site and nearest-neighbour Pauli terms. 
	The same core trends hold: the shallow decomposition consistently yields the highest trace distance and the largest variability across the ensemble, often exceeding the error of other methods by more than an order of magnitude. 
	In contrast, the wide decomposition again achieves the least trace distances, with minimal variation between instances.
	
\begin{figure}[ht!]
		\includegraphics[width=\linewidth]{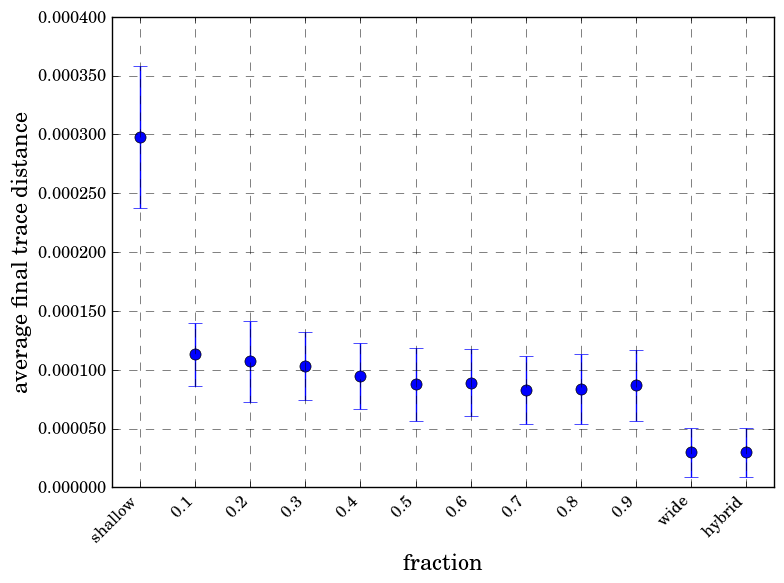}
		\captionsetup{width=\linewidth}
		\caption{Averaging the final value of the trace distance between the exact density matrix and the density matrix evolved using a Suzuki decomposition for an ensemble of 100 random Pauli Hamiltonians with on-site and nearest neighbour interactions. The horizontal axis shows the decomposition types: shallow, fractional for the wide:shallow fraction from 0.1 to 0.9, and wide/hybrid. \label{fig:average}}
	\end{figure}

	As shown in Fig. \ref{fig:average}, all fractional decompositions perform significantly better than the shallow decomposition, typically reducing the average trace distance by a factor of three or more, even for a wide fraction as low as 0.1. 
	No strong trend emerges as the wide fraction increases, indicating that the performance of fractional decompositions is relatively insensitive to the precise proportion used.
	Even if there is onlu a single wide decomposition step (this corresponds to a fraction of 0.1) is sufficient to escape the poor-performance regime of the shallow decomposition, consistently yielding a meaningful reduction in error across all three Hamiltonian families studied.
	This reinforces the observation that the intermediate region of decomposition space is surprisingly flat and populated by many relatively high-performing approximants.
	
	\subsubsection{Impact of Depolarising Noise on Decomposition Performance}
	In practical quantum computation, gate operations are subject to noise, and longer circuits are inherently more vulnerable to accumulated error. 
	To study how decomposition structure interacts with noise, we introduce a simple model of depolarising noise, applying a fixed probability of random error after each gate (i.e., after each exponential in the decomposition sequence). 
	Depolarising noise is a standard model for quantum errors, where the system is replaced by the maximally mixed state with probability $p$ and remains unchanged otherwise. 
	For a single application, the action of a depolarising channel $\mathcal{E}_p$ on a density matrix $\rho$ is given by
	\begin{align} 
		\mathcal{E}_p(\rho) = (1-p)\rho + p\frac{\textnormal{I}}{d}, 
	\end{align} 
	where $d$ is the dimension of the Hilbert space.
	
	This setup allows us to investigate how the trade-off between decomposition error and gate-induced noise shifts as a function of the physical error rate $p$. 
	In particular, we expect crossover behaviour: decompositions with lower intrinsic approximation error but longer sequence lengths (e.g., wide) may become less advantageous as $p$ increases, compared to shorter decompositions (e.g., shallow) which accumulate less noise but have higher approximation error.
	
\begin{figure}[ht!]
		\includegraphics[width=\linewidth]{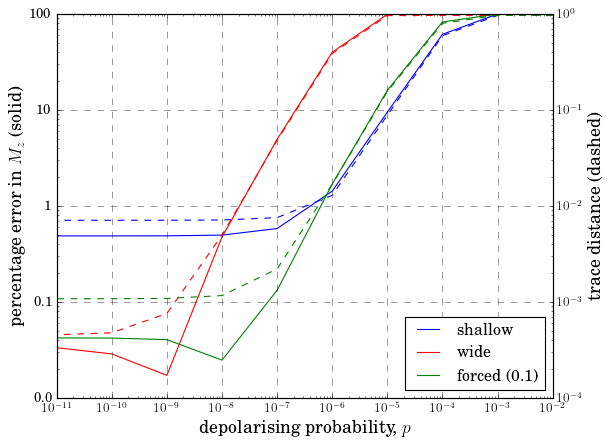}
		\captionsetup{width=\linewidth}
		\caption{Effect of depolarising noise on decomposition accuracy for the TFIM. Solid lines show the percentage error in final magnetization; dashed lines show the trace distance to the exact density matrix. As the depolarising probability $p$ decreases, the optimal decomposition shifts from shallow (shortest) to fractional (intermediate), with wide (longest) only outperforming others at unphysically low noise rates. \label{fig:noise}}
	\end{figure}

	In Fig. \ref{fig:noise}, we study this behaviour using the transverse field Ising model (TFIM) and three representative decompositions: the shallow decomposition (standard Strang splitting), the wide decomposition (maximal sequence length), and a fractional decomposition with a single wide step (wide:shallow ratio of 0.1).
	The left vertical axis shows the percentage error in the final magnetization, and the right vertical axis shows the final trace distance between the exact and noisy density matrices.
	Sweeping the depolarising probability from $10^{-11}$ to $10^{-2}$, we observe that the optimal decomposition strategy depends strongly on the noise strength.
	This range spans both the noisy intermediate-scale quantum (NISQ) regime, where two-qubit gate error rates are typically between $10^{-2}$ and $10^{-3}$, and the regime anticipated for scalable fault-tolerant quantum computation, where error rates are targeted around $10^{-4}$ for two-qubit gates and $10^{-6}$ for single-qubit gates, into the fully fault-tolerant regime for even smaller $p$.
	
	For depolarising probabilities $p > 10^{-6}$, simulation error is dominated by noise accumulation, and the shallow decomposition achieves the lowest total error due to its minimal sequence length.
	As $p$ decreases below $10^{-6}$, decomposition error begins to dominate, and the fractional decomposition with a single wide step outperforms the shallow strategy, achieving a final magnetization error below 0.1\% while maintaining a significantly smaller gate count than the wide decomposition.
	The wide decomposition, despite its lower intrinsic approximation error, only becomes advantageous at unphysically small noise rates $p < 10^{-9}$, where the cumulative effect of noise becomes negligible.
	These results show that in near-term quantum hardware, where gate errors are expected to reach $10^{-4}$ to $10^{-5}$, the shallow decomposition remains preferable.
	However, in future fault-tolerant regimes with gate errors approaching $10^{-6}$ or below, fractional decompositions offer a practical balance between noise resistance and approximation accuracy, providing an attractive alternative to shallow and wide constructions.
	
	\section{Discussion}
	In this work, we have explored the landscape of second-order Suzuki product formula decompositions, focusing on the flexibility and underexamined richness of their structure when applied recursively to Hamiltonians composed of multiple non-commuting terms.
	While standard Strang splitting is commonly used, it is only one of many possible decompositions that share the same theoretical error bounds but differ in sequence length, gate count, and simulation performance.
	By considering sequential applications of the Suzuki formula, we showed how a wide range of valid approximants can be generated through different combinations of term orderings and decomposition patterns, bounded between the minimal-depth Strang form and a binary-tree-like maximum-depth construction.
	
	To navigate this combinatorial space, we proposed a hybrid heuristic that selects at each step the decomposition (shallow or wide) that minimizes the local error bound to try and balance accuracy and gate efficiency.
	We also introduced a parameterized scheme, the fractional decomposition, that fixes the proportion of wide decomposition steps to systematically explore the trade-off between structure and performance.
	These strategies offer tractable ways to probe the impact of decomposition structure without resorting to intractable global optimization.
	The empirical behaviour of these approximants was then benchmarked across three representative 1D spin models.
	
	The shallow decomposition consistently yields the largest simulation error and greatest variability across all models, despite having the shortest sequence length and being the most widely used.
	The wide decomposition achieves the lowest error, but at a substantial cost in gate count.
	The hybrid decomposition, which minimizes the local error bound at each recursive step, always recovers the wide decomposition, suggesting that it is locally optimal under current analytic bounds.
	
	Fractional decompositions provide a systematic way to interpolate between shallow and wide structures.
	By fixing the proportion of wide steps in advance, they offer a simple and compiler-friendly strategy that can be evaluated offline.
	In all models considered, even a single wide step (fraction 0.1) is sufficient to improve accuracy significantly over the shallow decomposition, often approaching the performance of the wide strategy.
	In some cases, fractional decompositions with substantially lower gate counts outperform the wide decomposition, demonstrating that global accuracy does not necessarily correlate with the local error bounds used in the heuristic.
	This highlights a limitation of the hybrid decomposition: local error minimization does not necessarily yield the best global approximant in terms of actual simulation error, and that the intermediate regime between shallow and wide approximates likely contains more viable approximants
	
	These results support two practical heuristics for simulation design.
	If minimizing decomposition error is critical, a sweep over fractional decompositions can be performed offline to identify low-error sequences with reduced gate cost, sometimes exceeding the performance of the hybrid decomposition.
	Otherwise, introducing a single wide decomposition step into the shallow structure yields an order-of-magnitude improvement in simulation error for only a modest increase in gate count.
	When depolarising gate noise is included, these heuristics remain robust.
	At current and near-term noise levels, the standard shallow decomposition remains preferable, but as error rates approach the fault-tolerant regime, fractional decompositions with a single wide step offer a favorable balance between noise accumulation and approximation accuracy.
	
	Two key questions remain open.
	Why does the shallow decomposition consistently perform poorly despite its symmetry and theoretical grounding?
	And why does the hybrid strategy always converge to the wide decomposition, even when better approximants exist?
	A more complete understanding of these questions will be important for developing efficient and generalizable decomposition heuristics.
	
	\acknowledgements
	This work was supported by the Engineering and Physical Sciences Research Council [grant numbers EP/T001062/1, EP/Y004140/1].
	We thank Chris Barrett, Aleks Kissenger, Noah Linden, Tom Melham  and Marcin Szyniszewski for stimulating and insightful discussions.
		
	\bibliographystyle{unsrtnat}
	\bibliography{document}
\end{document}